\documentclass[12pt]{article}
\usepackage{amsfonts,amsmath}
\usepackage[mathscr]{eucal}
\usepackage{amssymb}
\usepackage{amsthm}
\theoremstyle{plain}

\usepackage{epsfig,epsf}
\usepackage{graphicx, subfigure}

\newcommand{\be}{\begin{eqnarray}}
\newcommand{\ee}{\end{eqnarray}}

\newcommand{\lb}{\label}
\newcommand{\p}[1]{(\ref{#1})}

\begin{document}

\begin{titlepage}

\vspace*{0.2cm}

\renewcommand{\thefootnote}{\star}
\begin{center}

{\LARGE\bf  Exceptional points of infinite order give a continuous spectrum  }\\

\vspace{1.5cm}
\renewcommand{\thefootnote}{$\star$}

{\large\bf Andrei~Smilga} \vspace{0.5cm}

{\it SUBATECH, Universit\'e de Nantes,}\\
{\it 4 rue Alfred Kastler, BP 20722, Nantes 44307, France;}\\

\vspace{0.1cm}

{\tt smilga@subatech.in2p3.fr}\\

\end{center}
\vspace{0.2cm} \vskip 0.6truecm \nopagebreak

\begin{abstract}
The statement in the title discussed earlier in association with the Pais--Uhlenbeck oscillator with equal
frequencies is illustrated for an elementary matrix model. In the limit
$N \to \infty$ ($N$ being the order of the exceptional point), an infinity of nontrivial states that do not change their
norm during evolution appear. These states have  real  energies
 lying in a continuous interval.
The norm of the ``precursors'' of these states at large finite $N$ is not conserved,
but the characteristic time scale where this nonconservation shows up grows
linearly with $N$.
\end{abstract}

\vspace{6cm}

\hfill {\it Dedicated to the people of Novorossia}

\newpage

\end{titlepage}

\setcounter{footnote}{0}

\setcounter{equation}0

Exceptional points are points in the space of the parameters of the Hamiltonian where two or more
eigenstates coalesce such that the Hamiltonian involves a Jordan block at this point
\cite{Heiss1}.
Such a Hamiltonian is not Hermitian, and the corresponding evolution operator is not unitary. Our main remark
is that for an exceptional point of {\it infinite} order, the Hermiticity and unitarity may in some cases
be restored. The spectrum
of the Hamiltonian thus obtained is real and {\it continuous}.

This phenomenon was earlier observed and studied for the Pais--Uhlenbeck oscillator
\cite{PU}   with the Lagrangian
 \be
L_{PU} = \frac 12 (\ddot{q} + \Omega_1^2 q)(\ddot{q} + \Omega_2^2 q)
  \ee
One can observe that the point
 $\Omega_1 = \Omega_2$ is
 an exceptional point of infinite order \cite{jaPhysLett}
(an infinity
of eigenstates that were distinct at $\Omega_1 \neq \Omega_2$
 coalesce at this point forming  Jordan blocks of infinite size).
However, it was later shown  \cite{PUSigma}
that these infinite Jordan blocks do not lead in this case
to the loss of Hermiticity and
unitarity,  but
signal the appearance of a {\it continuous spectrum}.
\footnote{
 The fact that the Hamiltonian of the PU oscillator with equal
frequencies has a real continuous spectrum
(which, however, is not bounded from below, nor from above) was actually
known since the original Pais and Uhlenbeck paper. We can also refer the
 reader  to more recent Ref.\cite{Bolonek}
for a nice detailed analysis of this issue.}

To avoid possible confusion, we emphasize that the statement above refers
to the {\it conventional} PU oscillator. Bender and Mannheim
recently suggested an unconventional realization of this system \cite{BM1,BM2}.
In constrast to the standard
PU oscillator Hamiltonian that involves ghosts
(the states with arbitrarily low energies),  their
$PT$--symmetric Hamiltonian is positive definite.
When $\Omega_1 \neq \Omega_2$, both the standard PU Hamiltonian and
the BM Hamiltonian are Hermitian.
 In the singular equal-frequency limit, the dynamics of these two different
quantum problems
is essentially different. In this limit the BM Hamiltonian
does not involve  infinite Jordan blocks, but an infinite set
 of Jordan blocks, each of finite size. The latter {\it  break} Hermiticity
and unitarity \cite{PUSigma} (we do not agree
 with  Ref.~\cite{BM2} on this issue).
On the other hand, for the conventional PU Hamiltonian,
 the presence of an  infinite number of coalescing states
does not lead to the breaking of unitarity.

The aim of this note is to clarify this phenomenon by studying
 a simple matrix model.
Consider the equation
  \be
\lb{eqmatr}
 i \frac {d\psi }{dt} \ =\ H \psi \, ,
  \ee
where $\psi = \left( \begin{array}{c} a \\ b \end{array} \right) $
is a two-component vector
and $H$ has the Jordan form
  \be
\lb{Ham2}
  H \ =\ \left( \begin{array} {cc} 0 & 1 \\ 0 & 0 \end{array} \right) \, .
  \ee
A general solution to the equation \p{eqmatr} is
 \be
 \lb{sol2}
 \psi(t) \ =\ \left( \begin{array}{c} a - ibt \\ b \end{array} \right)  \, .
 \ee
If $b \neq 0$, the (conventionally defined)
norm of the state \p{sol2} grows with time such that the evolution is not
unitary. For the subspace
$\psi = \left( \begin{array}{c} a \\ 0 \end{array} \right)$,
the evolution is unitary, however.

Consider now an exceptional point of order $N$.
It can be described by the matrix Schr\"odinger equation \p{eqmatr} where
$\psi(t)$ is now an $N$-component vector and the Hamiltonian
represents the matrix \cite{Heiss2}
\footnote{A more complicated matrix model for multiple exceptional
points was earlier suggested
in Ref.\cite{Znojil}.}
 \be
\lb{HamN}
  H_N \ =\ \left( \begin{array} {cccccc} 0 & 1  & 0 & 0 & \cdots & 0 \\
0 & 0  & 1 & 0 & \cdots & 0 \\ 0 & 0  & 0 & 1 & \cdots & 0 \\
\cdots & \cdots  & \cdots & \cdots & \cdots & 1
\\ 0 & \cdots  & \cdots & \cdots & \cdots & 0
 \end{array} \right) \, .
  \ee
A general solution to Eq. \p{eqmatr} with the Hamiltonian \p{HamN} with
the initial conditions
  \be
\lb{psi_aj}
 \psi(0) \ =\ \left( \begin{array} {c} a_0 \\ a_1 \\ \cdots \\ a_{N-1}
\end{array} \right)
  \ee
is
  \be
\lb{solN}
   \psi(t) \ =\ \left( \begin{array}{c} a_0 - ia_1t
- \frac { t^2}2 a_2 + \cdots  + \frac {(-it)^{N-1}}{(N-1)!} a_{N-1}  \\
a_1 - ia_2t
- \frac { t^2}2 a_3 + \cdots  + \frac {(-it)^{N-2}}{(N-2)!} a_{N-1} \\
\cdots  \\ a_{N-2} - i a_{N-1} t \\ a_{N-1}
 \end{array} \right)  \, .
  \ee
The evolution is unitary only if $a_1 = \ldots = a_{N-1} = 0$. But,
generically, it is not.

Let us, however, impose  the special initial conditions,
 \be
\lb{basis_eps}
a_j \ =\ \epsilon^j
  \ee
with a real $\epsilon$.
One can observe that in this case the components of the vector \p{solN} represent  truncated exponentials,
    \be
\lb{solNE}
   \psi(t) \ =\ \left( \begin{array}{c} E_N(-i\epsilon t)  \\
\epsilon  E_{N-1}(-i\epsilon t) \\ \cdots \\ \epsilon^{N-1}
 \end{array} \right)  \, ,
  \ee
where $E_j(x) = \sum_{k=0}^{j-1}  \, \frac {x^k}{k!} $.
In the limit $N \to \infty$, the exponentials are no longer truncated and the solution
\p{solNE} goes over to
   \be
\lb{solinfty}
   \psi(t) \ =\ \left( \begin{array}{c} 1  \\
\epsilon  \\ \cdots \end{array}
 \right) e^{-it \epsilon } \, .
  \ee
If $|\epsilon| < 1$, the norm of this state,
$$
\| \psi(t) \|^2 \ =\ \sum_{j=0}^\infty \, \epsilon^{2j} \, = \ \frac 1{1-\epsilon^2} ,$$
is finite
 and does not depend on time. We can multiply the wave function \p{solinfty} by $\sqrt{1-\epsilon^2}$ to normalize it.
Obviously, the parameter
$\epsilon$ (it can either be positive or negative) has the meaning of energy.

We have derived that, whereas, at any finite $N$, the evolution is unitary only in a limited 1-dimensional subspace with
only one nonvanishing component of the vector $\psi$, for $N \to \infty$, an infinity of certain special
state vectors appear whose norm does not grow with time.
 It is instructive to explore how this limit is achieved.
The norm of the finite  $N$ states \p{solNE} is
   \be
  \lb{norm}
  \|\psi(t)\|^2 \ =\ \sum_{j=1}^N \, \epsilon^{2(N-j)} |E_j(-i \epsilon t)|^2
   \ee
 If $t$ is not too large, the R.H.S. of Eq.~\p{norm} is
roughly constant. An elementary qualitative analysis shows
that the norm becomes a nontrivial ``live''  function of time
at the characteristic scale
  \be
\lb{tchar}
t^* \ \sim \ \frac N{C\epsilon} \, ,
  \ee
with $C \sim 1$. Numerical estimates show that $C \approx 2.5$.
We present in  Fig.~1 the time dependence of $\|\psi(t) \|$ for
$ N = 10$ and $N= 20$ with energy $ \epsilon = 0.2 $.

\begin{figure}[ht!]
     \begin{center}

        \subfigure[ N = 10 ]{
            \label{hren1}
            \includegraphics[width=0.4\textwidth]{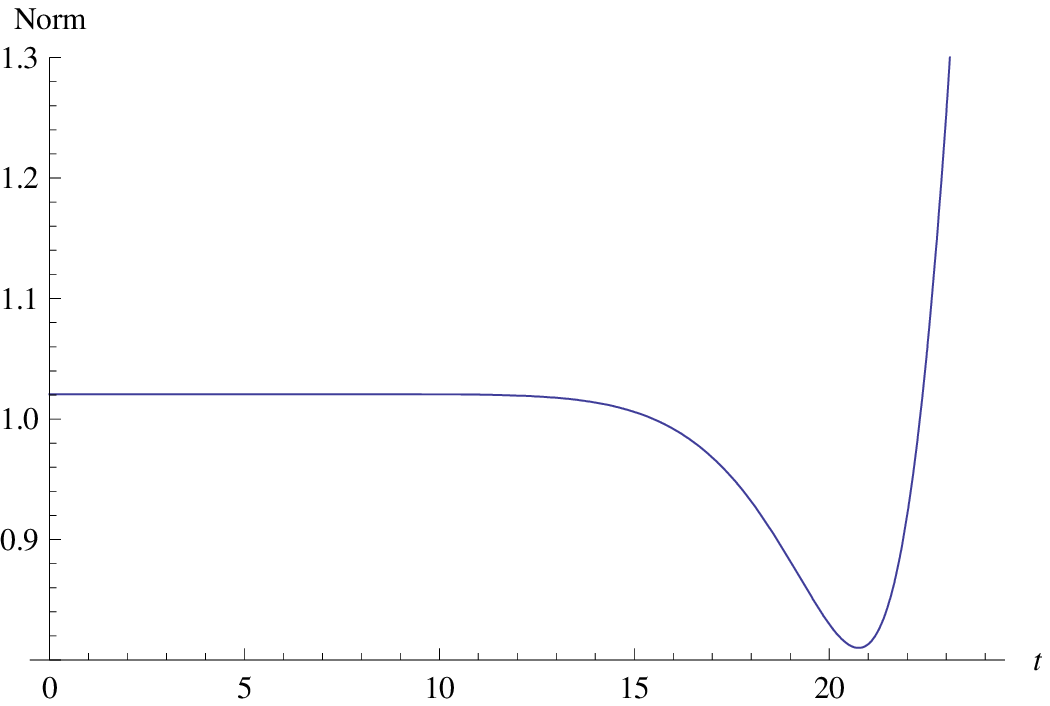}
        }
        \subfigure[N = 20 ]{
           \label{hren2}
           \includegraphics[width=0.4\textwidth]{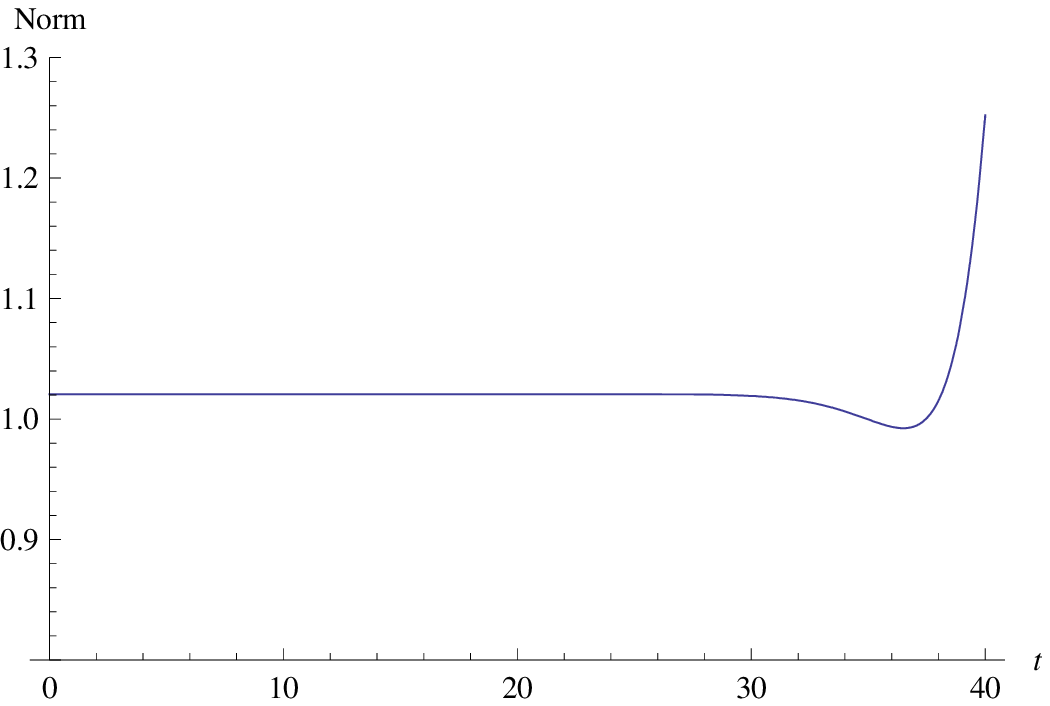}
        }
    \end{center}
    \caption{ Norm $\|\psi \|$ of the states \p{solNE} with $\epsilon = 0.2$ as a function of time. }
    \end{figure}

As far as the $N \to \infty$ limit of the described matrix model is concerned, one can make the following observations:

\begin{enumerate}
\item
  The states \p{basis_eps} represent a complete basis in the space of normalized vectors
\p{psi_aj}.
 Indeed, consider
a superposition of the states \p{basis_eps} with the weight function $f(\epsilon)$.  The conditions
  \be
\lb{int_eps_a}
 \int_{-1}^1 f(\epsilon) \epsilon^j d\epsilon \ =\  a_j
 \ee
can be resolved by orthogonalizing \p{int_eps_a} and representing  $f(\epsilon)$ as a series over Legendre
polynomials $P_n(\epsilon)$.

\item
The existence of the stationary--norm solutions does not mean here that the evolution is unitary for all initial
conditions. Indeed, if we start at $t=0$ with the vector $a_j = \epsilon^j + \mu^j$, $\epsilon \neq \mu$,
its norm will depend
on time. This is a corollary of the fact that the different states \p{basis_eps} are not mutually
orthogonal.

\item
One can also construct a state whose norm grows exponentially with time.
 Indeed, impose the initial conditions
  \be
\lb{alpha}
a_j \ =\ \alpha^j
  \ee
  with complex $\alpha$, $|\alpha| < 1, \ {\rm Im}(\alpha) > 0$.
This is an eigenfunction of the Hamiltonian $H^* \ =\ \lim_{N \to \infty} H_N$ with $H_N$ defined in \p{HamN}.
\footnote{In this limit, the Hamiltonian \p{HamN} coincides after a similarity transformation with the annihilation
operator of a harmonic oscillator,
 \be
\lb{annil}
a \ =\ \left( \begin{array} {cccccc} 0 & 1  & 0 & 0 & \cdots & 0 \\
0 & 0  & \sqrt{2} & 0 & \cdots & 0 \\ 0 & 0  & 0 & \sqrt{3} & \cdots & 0 \\
\cdots & \cdots  & \cdots & \cdots & \cdots & \cdots
 \end{array} \right) \, .
   \ee
The eigenstates \p{alpha} are then nothing but coherent states. To avoid confusion, bear in mind, however, that,
unlike the annihilation operator \p{annil}, the Hamiltonian \p{HamN} is responsible, as any Hamiltonian is,
for the time evolution of the system.}

Its time evolution boils down to multiplying by the
exponentially growing factor $e^{-i\alpha t}$. One can interpret $\alpha$ as
 ``complex energy''.
 Note that the state \p{alpha} can, as any other state, be represented as a superposition of the basis states
\p{basis_eps}. The corresponding weight function is
  \be
f(\epsilon) \ =\ \frac 12
\sum_{n=0}^\infty (2n+1)\, P_n(\epsilon) P_n(\alpha) \, .
  \ee

\item Nonconservation of the norm for generic initial conditions
 means that the evolution operator is not unitary and the Hamiltonian is
not Hermitian.

\item One can still ask whether the Hamiltonian $H^*$ might
 belong to the class
of pseudo-Hermitian or, better to say,
crypto-Hermitian
Hamiltonians \cite{crypto}. Crypto-Hermitian Hamiltonians  are  Hamiltonians that
are not Hermitian, but have a real spectrum.
By changing the definition of the norm, they can be rendered manifestly Hermitian.

In our case, the Hamiltonian has eigenvectors with real eigenvalues which constitute a complete basis
in the Hilbert space. One might hope to unravel the crypto-Hermitian structure of $H^*$
  by constructing  an approximation to it by finite matrices  (cf. \cite{Levai})
which are {\it different} from the original matrix Hamiltonians \p{HamN} (the latter are  evidently not Hermitian, nor
  crypto-Hermitian) and are crypto-Hermitian at any finite $N$. For any finite $N$, one could then  perform
a similarity transformation associated with the norm redefinition  making the states orthogonal and the Hamiltonian
Hermitian. We
attempted to pursue this program in the Appendix. We found out that it {\it does not} work --- as $N$ grows, the system
of $N$ eigenvectors thus constructed becomes less and less independent, the determinant of their components vanishing
exponentially fast.

Thus, even though one can construct finite-N crypto-Hermitian matrix approximants, the limit $N \to \infty$ is singular.

 \end{enumerate}

On the other hand, for the PU oscillator with equal frequencies, we have an ordinary continuous spectrum with
a unitary evolution operator. An essential difference of this system from the matrix model
studied in this paper
is the fact that the Jordan block structure appears there in the basis of ``bad'' unnormalizable states. (When $\Omega_1 \neq \Omega_2$, the corresponding states are quite ``good'', belonging to ${\cal L}_2$. But in the equal-frequency limit, the exponential factor rendering them normalizable disappears.)

We refer the reader to Refs.\cite{jaPhysLett,PUSigma} for a more detailed
 analysis of this system and only discuss here,
following \cite{PUSigma}, a trivial model with the same physics.
Consider  the  Hamiltonian describing free 1-dimensional motion,
   \be
\lb{Hfree}
H \ =\ - \frac 12 \frac {\partial^2}{\partial x^2}
    \ee
It has the continuous spectrum eigenfunctions
 \be
\label{Psifree}
 \Psi_k(x; t) \ =\ \exp\left\{ ikx - \frac {ik^2}2 t \right\}\ .
 \ee
Note now that not only (\ref{Psifree}), but also every term of its expansion
in $k$,
 \be
\label{kuski}
k^0\ :\ \ \Psi_0(x;t) &=& 1 \, , \nonumber \\
k^1\ :\ \ \Psi_1(x;t) &=& x \, , \nonumber \\
k^2\ :\ \ \Psi_2(x;t) &=& t - ix^2 \, , \nonumber \\
k^3\ :\ \ \Psi_3(x;t) &=& xt - \frac {ix^3}3 \, , \nonumber \\
k^4\ :\ \ \Psi_4(x;t) &=& \frac{t^2}2 - itx^2 - \frac {x^4}6  \, ,
 \ee
etc.,  satisfies the time-dependent Schr\"odinger equation \p{eqmatr}
with the Hamiltonian \p{Hfree}.
Obviously, the functions (\ref{kuski}) are not normalizable. Still, one can observe that the Hamiltonian \p{Hfree}
expressed in this unusual basis acquires a nondiagonal Jordan--like form,
\footnote{The Hilbert space of the Hamiltonian \p{Hfree} has two sectors with the functions even and odd under $x \to -x$.
 Correspondingly,
we have here two Jordan ladders.}

  \be
H \Psi_0 =  H \Psi_1 = 0,\ \ H\Psi_2 = i \Psi_0, \ \ H \Psi_3 = i \Psi_1, \ \
H \Psi_4 =  i\Psi_2, \ {\rm etc.}\, ,
  \ee
 and one can relate the continuity of the spectrum of \p{Hfree} to this fact. ``Bad'' (not normalizable)
basis leads to ``good''  [mutually
orthogonal in the usual sense, $\int \, \Psi^*_{k'} \Psi_k \, dx \sim \delta(k'-k)$]
continuous spectrum eigenfunctions.

And, for the matrix model, ``good'' (normalizable) basis states
\p{basis_eps} result in  ``bad''  continuous spectrum states --- 
nonorthogonal real-energy states 
and also states with complex energies.

I am indebted to W.D. Heiss for useful discussions and to U. Guenther, H.F. Jones  and M. Znojil for many
illuminating discussions and valuable comments.

\section*{Appendix}
\setcounter{equation}0
\def\theequation{A.\arabic{equation}}

\,

We explain here why an attempt to construct a
sequence of crypto-Hermitian matrix Hamiltonians with the same limit $N \to \infty$ as the limit
$H^*$ of the Hamiltonians \p{HamN} fails.

Consider the normalized eigenstates of $H^*$,  $|\epsilon \rangle: \ a_j = \sqrt{1-\epsilon^2}  \, \epsilon^j $ with real
$\epsilon \in (-1,1)$. We have
 \be
\lb{epsmu}
 \langle \epsilon|\mu \rangle \ =\ \frac {\sqrt{(1-\epsilon^2)(1-\mu^2)}}{1-\epsilon \mu} \, .
  \ee
Consider now a set of $N = 2M+1$ $N$-dimensional unitary vectors
$|n = -M, \ldots,0,\ldots, M\rangle$ with the inner products
  \be
\lb{amn}
 \langle n | m \rangle \ =\ \frac {\sqrt{(1-\epsilon_n^2)(1-\epsilon_m^2)}}{1 - \epsilon_n \epsilon_m},
\ \ \ \ \ \ \ \ \ \ \epsilon_n = \frac n{M+1} \, .
  \ee
For example, for $M=1$, we choose
  \be
\lb{vectory}
 |- \rangle \ =\ \left( \begin{array}{c} \frac {\sqrt{3}}2 \\ \frac 1{\sqrt{20}} \\  - \frac 1{\sqrt{5}} \end{array} \right),
\ \ \ |0 \rangle \ =\ \left( \begin{array}{c} 1 \\ 0 \\ 0 \end{array} \right), \ \ \ \
 |+ \rangle \ =\ \left( \begin{array}{c} \frac {\sqrt{3}}2 \\ \frac 1{\sqrt{20}} \\
\frac 1{\sqrt{5}} \end{array} \right) \, .
  \ee
Obviously, \p{amn} represents a discretization of \p{epsmu}.

At the next step, we construct the matrix $(\tilde H)_N$ having the vectors $|n \rangle$ as eigenvectors with eigenvalues
$\epsilon_n$. This is possible to do as long as the system of $N^2$ equations
 \be
\lb{eqforH}
 ({\tilde H}_N)_{ij} a^{(n)}_j \ =\ \epsilon_n a^{(n)}_i
 \ee
for the matrix elements $   ({\tilde H}_N)_{ij}$ is not degenerate. In other words, as long as
the vectors $|n\rangle$ are linearly independent and the determinant of the $N \times N$ matrix made of
the vector components does not vanish. For $M=1$, this determinant is equal to 0.2. This indicates that, though the vectors
\p{vectory} are not coplanar, they are relatively close to being so. Finally, one can make these vectors orthogonal  by redefining
the norm in an appropriate way.

However, the larger  $M$ and $N$ are, the more difficult is to carry on this program.
It is more convenient to study not the determinant of the eigenvector components,
but its square, the Gram determinant of the matrix of their
scalar products.
For $M=1$, the Gram deteminant $\Delta_1$ of the
scalar products \p{amn} is equal to 0.04. For $M=2$, it is equal to $10^{-5}$, meaning that the
corresponding five eigenvectors are ``almost'' linearly dependent. For $M=3$, it is already $10^{-11}$. It roughly decays as
$\Delta_M \sim e^{-1.2M^2}$.
\footnote{To rigorously justify this numerical observation is an interesting problem for a mathematics student.}
A matrix with an almost degenerate system of eigenvectors must have very large elements. The limit $N \to \infty$ is
singular.

\end{document}